\documentclass[12pt]{article}
\usepackage{amsmath}
\usepackage{amsfonts}        
\usepackage{amssymb}
\usepackage{amsbsy}
\usepackage{mathrsfs}
\usepackage{epsfig}      
\usepackage{color}       

\usepackage[T1]{fontenc} 

\usepackage{graphicx}

\newlength{\TZ}
\DeclareFontFamily{OT1}{pzc}{}
\DeclareFontShape{OT1}{pzc}{m}{it}{<-> s * [1.200] pzcmi7t}{}
\DeclareMathAlphabet{\mathpzc}{OT1}{pzc}{m}{it}
\newcommand{\BEQ}{\begin{equation}}     
\newcommand{\BEA}{\begin{eqnarray}}
\newcommand{\BD}{\begin{displaymath}}
\newcommand{\EEQ}{\end{equation}}       
\newcommand{\EEA}{\end{eqnarray}}
\newcommand{\ED}{\end{displaymath}}

\newcommand{\vep}{\varepsilon}          
\newcommand{\D}{{\rm d}}                
\newcommand{\erfc}{{\rm erfc\,}}        
\newcommand{\demi}{\frac{1}{2}}         
\newcommand{\wit}[1]{\widetilde{#1}}    

\renewcommand{\vec}[1]{\boldsymbol{#1}} 

\newcommand{\appsektion}[1]{\setcounter{equation}{0}\setcounter{subsection}{0}
\section*{Appendix. #1}
\renewcommand{\theequation}{A.\arabic{equation}}
              \renewcommand{\thesection}{A}
              \renewcommand{\thefigure}{A\arabic{figure}}\setcounter{figure}{0} }

\catcode`\@=11
\def\numberbysection{\@addtoreset{equation}{section}
        \def\theequation{\thesection.\arabic{equation}}}
\numberbysection                        

\definecolor{gruen}{rgb}{0,0.625,0}       
\definecolor{rot}{rgb}{0.75,0,0}          
\definecolor{blau}{rgb}{0,0,0.75}         
\definecolor{casta}{rgb}{0.45,0.20,0}     
\definecolor{gelb}{rgb}{0.825,0.725,0.0}  

\begin{document}

\begin{titlepage}

\vskip 1.5 cm
\begin{center}
{\LARGE \bf Critical ageing correlators from Schr\"odinger-invariance}
\end{center}

\vskip 2.0 cm
\centerline{{\bf Malte Henkel}$^{a,b}$ and {\bf Stoimen Stoimenov}$^c$}
\vskip 0.5 cm
\centerline{$^a$Laboratoire de Physique et Chimie Th\'eoriques (CNRS UMR 7019),}
\centerline{Universit\'e de Lorraine Nancy, B.P. 70239, F -- 54506 Vand{\oe}uvre l\`es Nancy Cedex, France}
\vspace{0.5cm}
\centerline{$^b$Centro de F\'{i}sica Te\'{o}rica e Computacional, Universidade de Lisboa,}
\centerline{Campo Grande, P -- 1749-016 Lisboa, Portugal}
\vspace{0.5cm}
\centerline{$^c$ Institute of Nuclear Research and Nuclear Energy, Bulgarian Academy of Sciences,}
\centerline{72 Tsarigradsko chaussee, Blvd., BG -- 1784 Sofia, Bulgaria}
\vspace{0.5cm}

\begin{abstract}
For ageing systems, quenched onto a critical temperature $T=T_c$ such that the dominant noise comes from the thermal bath, with a 
non-conserved order-parameter and in addition with dynamical exponent $\mathpzc{z}=2$, the form of the 
two-time auto-correlator as well as the time-space form of the single-time correlator are derived from Schr\"odinger-invariance, generalised 
to non-equilibrium ageing. These findings reproduce the exact results in the $1D$ 
Glauber-Ising model at $T=0$ and the critical spherical model in $d>2$ dimensions. 
\end{abstract}
\end{titlepage}

\setcounter{footnote}{0}



\section{Introduction} \label{sec:1}

Physical ageing is an often-encountered example of collective phenomena in the non-equilibrium dynamics of many-body systems \cite{Stru78,Cugl03,Giam16,Vinc24}. 
In classical systems this is often brought about via a temperature-quench, from an initially highly disordered state, either directly onto the critical temperature $T=T_0$ 
\cite{Cugl03,Cala05,Puri09,Henk10,Taeu14} which will be in the focus of this work or else into the phase-coexistence region where 
$T<T_c$ \cite{Bray94a,Cugl03,Puri09,Henk10,Cugl15}. 
Since microscopically a rapid relaxation to a single equilibrium state is then impossible and the system becomes spatially inhomogeneous, physical ageing occurs. 
Phenomenologically, this may be characterised by the following three defining properties \cite{Stru78,Henk10}. 

\noindent
{\bf Definition:} {\it A many-body system is said to undergo {\em physical ageing} when its relaxation dynamics obeys
the properties
\begin{enumerate}
\item slow relaxation dynamics (becoming more slow with increasing time) 
\item absence of time-translation-invariance
\item dynamical scaling 
\end{enumerate}
}
\noindent 
All three properties are required to specify the physical phenomenon 
we have in mind. 
If one describes the relaxation process in terms of a (coarse-grained) {\em order-parameter} $\phi(t,\vec{r})$ and the initial state is
such that the average $\bigl\langle\phi_{\rm init}(\vec{r})\bigr\rangle=\bigl\langle \phi(0,\vec{r})\bigr\rangle=0$, 
it follows that $\langle \phi(t,\vec{r})\rangle=0$ for all times $t>0$.
Ageing is conveniently studied through the {\em two-time correlator} $C$ and the {\em two-time response} $R$, defined as 
\BEQ \label{gl:1}
C(t,s;{r}) = \bigl\langle \phi(t,\vec{r})\phi(s,\vec{0})\bigr\rangle \;\; , \;\;
R(t,s;{r}) = \left. \frac{\delta \bigl\langle \phi(t,\vec{r})\bigr\rangle}{\delta h(s,\vec{0})}\right|_{h=0} 
=\bigl\langle \phi(t,\vec{r})\wit{\phi}(s,\vec{0})\bigr\rangle
\EEQ
where the average is both over initial states as well as over thermal histories.
Spatial translation- and rotation-invariance such that $\vec{r}\mapsto r = |\vec{r}|$ are admitted for notational simplicity. 
We also anticipate from Janssen-de Dominicis theory \cite{Domi76,Jans76} rewriting $R$ with the so-called {\em response scaling operator} $\wit{\phi}(t,\vec{r})$. 
Single-time correlators $C(s;r) := C(s,s;r)$ are included by letting $t=s$. Auto-correlators $C(t,s):=C(t,s;\vec{0})$ and auto-responses 
$R(t,s):=R(t,s;\vec{0})$ are obtained from (\ref{gl:1}) for $r=0$.

\begin{figure}[tb]  
\includegraphics[width=.98\hsize]{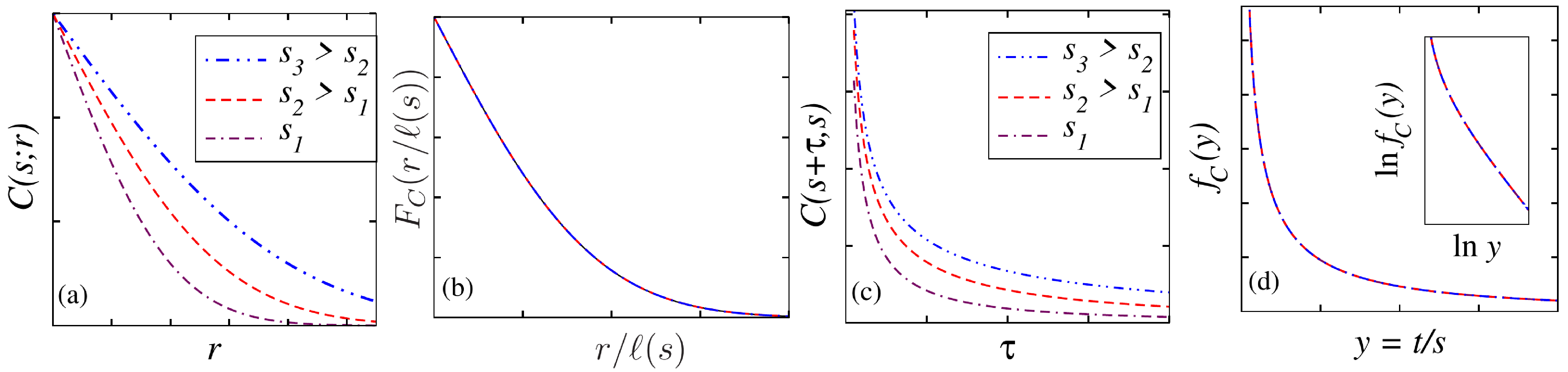}  
\caption[fig2]{Illustration of physical ageing in single-time and two-time correlators. 
A typical single-time correlator $C(s;r)$ is in panel (a) for different times
$s_1<s_2<s_3$ but collapses in panel (b), when replotted over against rescaled lengths $r/\ell(s)$. The dynamical length scale is $\ell(s)\sim s^{1/\mathpzc{z}}$.  
In panel (c) a typical two-time auto-correlator $C(s+\tau,s)$ is displayed over against $\tau=t-s$, for different waiting times $s_1<s_2<s_3$ 
which collapse when replotted in panel (d) over against $y=t/s$. The inset shows the asymptotic power-law form $f_C(y)\sim y^{-\lambda_C/z}$.
\label{fig2} }
\end{figure}

Figure~\ref{fig2} illustrates the defining properties of physical ageing after a quench into $T<T_c$. 
Clearly, the single-time correlator $C(s;r)$ evolves more slowly with $r$ when $s$ increases and
there are different curves for different values of $s$, see figure~\ref{fig2}a. The data collapse of dynamical scaling is shown in figure~\ref{fig2}b. 
Similarly, figure~\ref{fig2}c shows that the two-time correlator $C(s+\tau,s)$ evolve more slowly with $\tau$ when the waiting time $s$ increases and for each $s$ there is a different curve. 
The corresponding data collapse of dynamical scaling is shown in figure~\ref{fig2}d. Informations of this kind are formally cast into scaling forms, 
for sufficiently large times $t,s\gg \tau_{\rm micro}$ and $y=t/s>1$ ($\tau_{\rm micro}$ is a microscopic reference time)
\begin{subequations} \label{gl:intro} 
\BEQ \label{gl:2}
C(t,s;{r}) = s^{-b}   F_C\left( \frac{t}{s}; \frac{\bigl|\vec{r}\bigr|}{s^{1/\mathpzc{z}}}\right) \;\; , \;\;
R(t,s;{r}) = s^{-1-a} F_R\left( \frac{t}{s}; \frac{\bigl|\vec{r}\bigr|}{s^{1/\mathpzc{z}}}\right)
\EEQ
where $\mathpzc{z}$ is the {\em dynamical exponent} and $a,b$ are the {\em ageing exponents}. 
Their known values depend on whether $T<T_c$ or $T=T_c$. The scaling functions $F_{C,R}$ are universal. 
In particular, one has for $y\gg 1$ 
\BEQ \label{gl:3a}
f_C(y) = F_C(y,0) \sim y^{-\lambda_C/\mathpzc{z}} \;\; , \;\; f_R(y) = F_R(y,0) \sim y^{-\lambda_R/\mathpzc{z}}
\EEQ
where $\lambda_C$ is the {\em auto-correlation exponent} \cite{Huse89} and $\lambda_R$ is the {\em auto-response exponent}.  
It is expected, at least for short-ranged initial correlations, that
\BEQ \label{gl:lambda} 
\lambda_C = \lambda_R
\EEQ
The asymptotics (\ref{gl:3a}) of $f_C(y)$ for $y\gg 1$ is shown in figure~\ref{fig2}d. 
For a quench onto the critical point $T=T_c$, when the system undergoes non-equilibrium critical dynamics \cite{Godr02}, the behaviour is qualitatively
similar, from from (\ref{gl:2}) it is apparent that the raw data for $C$ (and analogously for $R$) must be multiplied by a convenient power of $s$ in order
to obtain dynamical scaling. When this is done, eqs.~(\ref{gl:3a},\ref{gl:lambda}) hold as well at $T=T_c$. 
How can one obtain these universal scaling functions $F_{C,R}$ in a generic way, without treating specific models explicitly, one after the other~? 
\end{subequations}

This work is organised as follows. Throughout, we shall restrict to $\mathpzc{z}=2$. 
Section~2 contains background on the non-equilibrium Janssen-de Dominicis field theory(specifically for non-equilibrium critical dynamics), 
equilibrium Schr\"odinger-invariance and
its adaptation to a far-from-equilibrium setting. 
Section~3 presents first the generic calculations for the two-time auto-correlator and the single-time time-space correlator. 
Then these are compared to the exactly known results in the $1D$ Glauber-Ising model at $T=0$ and the spherical model at $T=T_c$ and $d>2$. We conclude in section~4. 
An appendix recalls properties of the Appell function $F_1$.

\section{Background} \label{sec:2}

\subsection{Non-equilibrium field-theory} 

This work is inserted into the context of non-equilibrium continuum field-theory \cite{Domi76,Jans76,Jans92,Taeu14}. 
Implicitly, we shall always restrict to non-conserved dynamics (`model A') of the order-parameter $\phi$. 
The average of an observable $\mathscr{A}$ is found from the functional integral 
\BEQ \label{dynft}
\bigl\langle \mathscr{A}\bigr\rangle = \int \mathscr{D}\phi\mathscr{D}\wit{\phi}\; \mathscr{A}[\phi]\, e^{-{\cal J}[\phi,\wit{\phi}]}
\EEQ
which already includes probability conservation \cite{Taeu14}.For non-equilibrium critical dynamics, and with the effects of the heat bath being described by a
white noise with temperature $T$, the Janssen-de Dominicis action reads 
\begin{align} \label{actionJdD} 
{\cal J}[\phi,\wit{\phi}] &= \int \!\D t\D\vec{r}\: \left( \wit{\phi} \left( \partial_t - \Delta_{\vec{r}} - V'[\phi]\right)\phi - T \wit{\phi}^2 \right) 
\end{align}
with the interaction $V'[\phi]$, the spatial laplacian $\Delta_{\vec{r}}$ (and the usual re-scalings).
We shall not consider here any contribution from a noisy initial configuration which merely leads to corrections to scaling when they are spatially short-ranged \cite{Pico04}. 
The {\em deterministic action} ${\cal J}_0[\phi,\wit{\phi}]  = \lim_{T\to 0} {\cal J}[\phi,\wit{\phi}]$ refers to temperature $T=0$.  
One defines {\em deterministic averages} $\bigl\langle \cdot \bigr\rangle_0$ as
\BEQ \label{detave} 
\bigl\langle \mathscr{A}\bigr\rangle_0 = \int \mathscr{D}\phi\mathscr{D}\wit{\phi}\; \mathscr{A}[\phi]\, e^{-{\cal J}_0[\phi,\wit{\phi}]}
\EEQ
by replacing in (\ref{dynft}) the action ${\cal J}$ by its deterministic part ${\cal J}_0$. 
{}From causality considerations \cite{Jans92,Cala05,Taeu14} or Local-Scale-Invariance ({\sc lsi}) \cite{Pico04} one has the Barg\-man su\-per\-se\-lec\-tion rules
\BEQ \label{Bargman}
\left\langle \overbrace{~\phi \cdots \phi~}^{\mbox{\rm ~~$n$ times~~}} 
             \overbrace{ ~\wit{\phi} \cdots \wit{\phi}~}^{\mbox{\rm ~~$m$ times~~}}\right\rangle_0 \sim \delta_{n,m}
\EEQ
for the deterministic averages. Only observables built from an equal number of order-parameters $\phi$ 
and conjugate response operators $\wit{\phi}$ can have non-vanishing deterministic averages. 
Response functions as in (\ref{gl:1}) can now be found directly as deterministic averages.
However, correlators are obtained from higher-point response functions \cite{Pico04,Henk10}
\BEQ \label{corrFT}
C(t,s;r) 
= \bigl\langle \phi(t,\vec{r}+\vec{r}_0)\phi(s,\vec{r}_0) \bigr\rangle 
= T \int_0^{\infty} \!\D u \int_{\mathbb{R}^d} \!\D\vec{R}\: \left\langle \phi(t,\vec{r}+\vec{r}_0) \phi(s,\vec{r}_0) \wit{\phi^2}(u,\vec{R}) \right\rangle_0
\EEQ
We shall see below that $\wit{\phi}_2 :=\wit{\phi^2}$ must be treated as a composite scaling operator. 
Eq.~(\ref{corrFT}) holds for non-equilibrium model-A-type dynamics, after a quench onto $T=T_c$. 
For the following symmetry characterisation we shall be obliged to restrict to systems where the dynamical exponent happens to be $\mathpzc{z}=2$.

\subsection{Schr\"odinger invariance} 

Calculations using (\ref{actionJdD},\ref{detave}) share many aspects of equilibrium calculations. 
The deterministic action ${\cal J}_0$ has the Schr\"odinger group as a dynamical symmetry \cite{Henk03a}. 
The Lie algebra is spanned by $\langle X_{\pm 1,0}, Y_{\pm\frac{1}{2}}, M_0\rangle$ (for simplicity in a $1D$ notation) where
\BEA
X_n &=& - t^{n+1}\partial_t - \frac{n+1}{2} t^n r\partial_r - (n+1) \delta t^n -\frac{n(n+1)}{4} {\cal M} t^{n-1} r^2 \nonumber \\
Y_m &=& - t^{m+\frac{1}{2}} \partial_r - \left( m + \frac{1}{2}\right) {\cal M} t^{m-\frac{1}{2}} \\
M_n &=& - t^n {\cal M} \nonumber
\EEA
which close into the Schr\"odinger Lie algebra $\mathfrak{sch}(1)$. 
This is the most large and finite-dimensional Lie algebra which sends each solution of $\bigl(2{\cal M}\partial_t - \partial_r^2\bigr)\phi=0$ into
another solution, as already known to Jacobi and to Lie, see \cite{Duva24} and references therein. 
In addition, $\delta$ is the scaling dimension and ${\cal M}>0$ the (non-relativistic) mass of the equilibrium field $\phi$. 
The form of the scaling generator $X_0$ implies the dynamical exponent $\mathpzc{z}=2$. 
The requirement of Schr\"odinger-covariance constrains the form of $n$-point deterministic averages or response functions \cite{Henk94}. 
We shall need  the two-point response function ($\mathscr{R}_0$ is a normalisation constant) 
\BEA
\lefteqn{\hspace{-2.5truecm}R(t_a,t_b;r) = \left\langle \phi_a(t_a,\vec{r}) \wit{\phi}_b(t_b,\vec{0})\right\rangle_0 
= \delta({\cal M}_a - {\cal M}_b)\, \mathscr{R}_0\, \delta_{\delta_a,\wit{\delta}_b}\: \Theta(t_a-t_b)} \nonumber \\ 
&\times&  \bigl( t_a - t_b\bigr)^{-2\delta_a} \exp\left[ - \frac{{\cal M}_a}{2} \frac{\vec{r}^2}{t_a-t_b} \right]
\label{2points}
\EEA
Response operators have negative masses such that $\wit{\cal M}_b=\wit{\cal M}=-{\cal M}=-{\cal M}_a<0$. The three-point response is
\BEA
\lefteqn{\hspace{-1.5truecm}\bigl\langle \phi_a(t_a,\vec{r}_a) \phi_b(t_b,\vec{r}_b)\wit{\phi}_c(t_c,\vec{r}_c)\bigr\rangle_0 = 
\delta({\cal M}_a + {\cal M}_b - {\cal M}_c) \Theta(t_{ac}) \Theta(t_{bc})\: }
\nonumber \\
&\times& t_{ac}^{-\delta_{ac,b}} t_{bc}^{-\delta_{bc,a}} t_{ab}^{-\delta_{ab,c}} \: 
\exp\left[ -\frac{{\cal M}_a}{2} \frac{\vec{r}_{ac}^2}{t_{ac}} - \frac{{\cal M}_b}{2} \frac{\vec{r}_{bc}^2}{t_{bc}}\right] 
\Phi_{ab,c}\left( \frac{\bigl[ \vec{r}_{ac}^2 t_{bc} - \vec{r}_{bc}^2 t_{ac} \bigr]^2}{t_{ab} t_{ac} t_{bc} }\right)
\label{3points}
\EEA
where we used that from (\ref{2points}) it follows $\wit{\delta}_c=\delta_c$ and with the shorthands
\BD
t_{ij} = t_i - t_j \;\; , \;\; \vec{r}_{ij} = \vec{r}_i - \vec{r}_j \;\; , \;\; \delta_{ij,k} = \delta_i + \delta_j - \delta_k
\ED
The Heaviside functions $\Theta$ take the causality constraints in (\ref{2points},\ref{3points}) into account \cite{Henk03a}. 
The function $\Phi_{ab,c}$ is not determined by Schr\"odinger-invariance. 

By construction, Schr\"odinger-invariance only applies to systems with a dynamical exponent $\mathpzc{z}=2$. However, this condition is only rarely met in
ageing systems quenched onto $T=T_c$, e.g. \cite{Cala05,Taeu14}, although we shall give some examples below. An eventual generalisation to
$\mathpzc{z}\neq 2$ would require a convincing identification of a physically sensible equation of motion and an analysis of the non-trivial
dynamical symmetries. While attempts to achieve this objective were made in the past, i.e. \cite{Henk02}, we do not consider these as physically
totally satisfactory \cite{Hinrichsen06a} and leave this problem as an open question for future work. 

\subsection{Far from equilibrium observables} 

Physical ageing occurs far from equilibrium and is not time-translation-invariant. 
Rather than dropping the time-translation generator $X_{-1}$ from the Schr\"odinger algebra, 
we use an inspiration from dynamical symmetries of non-equilibrium systems \cite{Stoi22} 
and propose to achieve far-from-equilibrium physics, without standard time-translation-invariance, 
by the following 

\noindent
{\bf Postulate:} \cite{Henk25c} {\it The Lie algebra generator $X_n^{\rm equi}$ of a time-space symmetry of an equilibrium system becomes a symmetry 
out-of-equilibrium by the change of representation}
\BEQ \label{gl:hyp}
X_n^{\rm equi} \mapsto X_n = e^{\xi \ln t} X_n^{\rm equi} e^{-\xi \ln t}
\EEQ
{\it where $\xi$ is a dimensionless parameter whose value contributes to characterise the scaling operator $\phi$ on which $X_n$ acts.} 

When applied to the dilatation generator $X_0$ this merely leads to a modified scaling dimension $\delta_{\rm eff} = \delta-\xi$, since 
\begin{subequations} \label{gl:Xgen}
\BEQ \label{gl:X0gen} 
X_0^{\rm equi} \mapsto X_0 = -t\partial_t - \frac{1}{\mathpzc{z}}r\partial_r - \bigl(\delta - \xi\bigr)
\EEQ
However, the time-translation generator $X_{-1}^{\rm equi}=-\partial_t$ now becomes
\BEQ \label{gl:X-1gen}
X_{-1}^{\rm equi} \mapsto X_{-1} = -\partial_t + \frac{\xi}{t}
\EEQ
\end{subequations}
Significantly, in this new representation the scaling operators become $\Phi(t) = t^{\xi} \phi(t) = e^{\xi \ln t}\phi(t)$ and the equilibrium response functions 
(\ref{2points},\ref{3points}) are adapted into non-equilibrium ones via 
(spatial arguments are suppressed for clarity)\footnote{This depends on the applicability of (\ref{gl:hyp}). Different changes of representation are conceivable \cite{Henk25c}.} 
\BEA
\bigl\langle \phi_a(t_a)\wit{\phi}_b(t_b)\bigr\rangle_0            &\mapsto & t_a^{\xi_a} t_b^{\wit{\xi}_b} \: \bigl\langle \phi_a(t_a)\wit{\phi}_b(t_b)\bigr\rangle_0 \nonumber \\
\bigl\langle \phi_a(t_a)\phi_b(t_b)\wit{\phi}_c(t_c)\bigr\rangle_0 &\mapsto & t_a^{\xi_a} t_b^{\xi_b} t_c^{\wit{\xi}_c} \: \bigl\langle \phi_a(t_a) \phi_b(t_b)\wit{\phi}_c(t_c)\bigr\rangle_0 
\label{reponseHE}
\EEA
This means that we now characterise a scaling operator $\phi$ by a pair of scaling dimensions $(\delta,\xi)$ and a response operator 
$\wit{\phi}$ by a pair $(\wit{\delta},\wit{\xi})$. 
As a consequence of Schr\"odinger-invariance (\ref{2points}), and the Bargman rule (\ref{Bargman}) with $n=m=1$, we have
\BEQ
\delta = \wit{\delta}
\EEQ
but $\xi$ and $\wit{\xi}$ remain independent. 

The above postulate was studied for two-time auto-responses and auto-correlators in \cite{Henk25,Henk25c} which explored in particular the consequences of requiring co-variance under
the two Lie generators $X_{-1,0}$ far from equilibrium. 
It was shown that it permits in particular to derive the generic properties (\ref{gl:intro}) of ageing, besides many more specific results. 
Here we want to test if the same idea also generalises to the three-point response (\ref{3points}) and shall use this to compute explicitly the scaling functions 
$F_C(1,u)$ and $f_C(y)$ of the single-time correlator and the two-time auto-correlator, see figure~\ref{fig2}. 
Such a model-independent calculation is only based on the dynamical symmetries.\footnote{Non-linearities, coming from the potential $V'[\phi]$, are irrelevant for large
times if $\xi>\demi$. This condition is satisfied in many systems at $T=T_c$ \cite{Henk25c}.}

\section{Calculation of correlators} \label{sec:3}

For the forthcoming comparison with specific models below, we recall first the result of combining (\ref{reponseHE},\ref{2points}) for the two-time response \cite{Henk25c} 
when $\mathpzc{z}=2$ 
\begin{subequations} \label{2reponse}
\begin{align} \label{2reponseR}
R(t,s;r) = \left\langle {\phi}(t,\vec{r}) {{\wit{\phi}}}(s,\vec{0})\right\rangle 
= \mathscr{R}_0 \, s^{-1-a} \left( \frac{t}{s}\right)^{1+a'-\lambda_R/2} 
\left(\frac{t}{s} -1 \right)^{-1-a'} \exp\left[ -\frac{\cal M}{2} \frac{r^2}{t-s} \right] 
\end{align}
where the exponents $a,a',\lambda_R$ of ageing are related to $\delta,\xi,\wit{\xi}$ as follows 
\BEQ \label{2reponse-exp}
\frac{\lambda_R}{2} = 2\delta -\xi \;\; , \;\; 1+a = 2\delta -\xi - \wit{\xi} \;\; , \;\; a'-a = \xi +\wit{\xi}
\EEQ
\end{subequations}
and ${\cal M},\mathscr{R}_0$ are non-universal, dimensionful constants. 

Our starting point is the combination of (\ref{corrFT}), valid for quenches onto $T=T_c$, 
wherein the three-point response $\langle\phi\phi\wit{\phi^2}\rangle_0$ is adapted to  non-equilibrium 
situations with (\ref{reponseHE}) and we also use the explicit form (\ref{3points}). 
Herein, the order parameter $\phi$ has the pair of scaling dimensions $(\delta,\xi)$, whereas for the composite response operator $\wit{\phi}_2 :=\wit{\phi^2}$ we have the pair 
$(2\wit{\delta}_2,2\wit{\xi}_2)$.\footnote{Non-trivial scaling dimensions for composite operators are a well-known fact 
at conformally invariant equilibrium phase transitions \cite{Zamo86,Itzy89,Anti25} and also in non-relativistic field-theories \cite{Moro11}.} 
Taking all this together leads to
\BEA
\lefteqn{C(t,s;r) = T \int_0^{\infty} \!\D u\: \Theta(t-u)\Theta(s-u)\, \bigl( ts\bigr)^{\xi} \bigl(t-s\bigr)^{-2(\delta-\wit{\delta}_2)} u^{2\wit{\xi}_2} 
\bigl(t-u\bigr)^{-2\wit{\delta}_2} \bigl(s-u\bigr)^{-2\wit{\delta}_2}}  \label{corr-gen} \\
&\hspace{-0.5truecm}\times&  \hspace{-0.5truecm}\int_{\mathbb{R}^d}\!\D\vec{R}\: 
\exp\left[ -\frac{\cal M}{2}\frac{(\vec{r}+\vec{r}_0-\vec{R})^2}{t-u}  -\frac{\cal M}{2}\frac{(\vec{r}_0-\vec{R})^2}{s-u} \right] 
\Phi\left( \frac{\bigl[ (\vec{r}+\vec{r}_0-\vec{R})(s-u) - (\vec{r}_0-\vec{R})(t-u)\bigr]^2}{(t-s)(t-u)(s-u)} \right)
\nonumber 
\EEA
with the un-determined scaling function $\Phi=\Phi_{\phi\phi,\wit{\phi^2}}$. 
Spatial translation-invariance is obvious since a shift in the integration variable $\vec{R}$ makes any
reference to the point $\vec{r}_0$ disappear.

\subsection{Two-time auto-correlator}

To evaluate $C(t,s)=C(t,s;0)$ we choose $t>s$ for definiteness and $r=0$. Changing variables, we have
\BEA
\lefteqn{ \hspace{-3.5truecm}C(t,s) = T \bigl( ts\bigr)^{\xi} \bigl(t-s\bigr)^{-2(\delta-\wit{\delta}_2)} 
\int_0^s \!\D u\: u^{2\wit{\xi}_2} \bigl(t-u\bigr)^{d/2-2\wit{\delta}_2} \bigl(s-u\bigr)^{d/2-2\wit{\delta}_2} \bigl(t+s-2u\bigr)^{-d/2} }\nonumber \\
&\times& \underbrace{~\int_{\mathbb{R}^d} \!\D\vec{P}\; e^{-\demi {\cal M} \vec{P}^2}\, \Phi\left( \vec{P}^2 \frac{t-s}{t+s-2u}\right)~}_{=:\, \Psi\bigl( (t-s)/(t+s-2u)\bigr)}
\EEA
since the integral on the last line defines a new function $\Psi$ of a single argument. We go over to a scaling description by letting $t=y s$ and $u=s w$ and find
\BEA
C(t,s) &=& T s^{-b} y^{d/2+\xi-2\wit{\delta}_2} \bigl(y-1\bigr)^{-2(\delta-\wit{\delta}_2)} \bigl(y+1\bigr)^{-d/2}  \label{3.4} \\
& & \times \int_0^1 \!\D w\: w^{2\wit{\xi}_2} \bigl(1-w\bigr)^{d/2-2\wit{\delta}_2} \bigl(1-y^{-1}w\bigr)^{d/2-2\wit{\delta}_2} \bigl(1-\frac{2}{y+1}w\bigr)^{-d/2}
\Psi\left(\frac{y-1}{y+1-2w}\right)
\nonumber
\EEA
Since the integral in the second line is finite in the $y\to\infty$ limit, we can read off the auto-correlation exponent $\lambda_C$ and the ageing exponent $b$
\BEQ \label{exposants}
\frac{\lambda_C}{2} = 2\delta - \xi \;\; , \;\; b = -2\xi +2\delta +2\wit{\delta}_2-2\wit{\xi}_2-1-\frac{d}{2}
\EEQ
The first identity is the same as in (\ref{2reponse-exp}) which proves the expectation (\ref{gl:lambda}). 
In the litt\'erature, this is derived either from field-theoretic methods \cite{Cala05,Maze04} 
or else from the combined dilatation- and generalised time-translation-invariance of the correlator and response \cite{Henk25,Henk25c}.
But it is very satisfying to see that (\ref{gl:lambda}) also follows from the Schr\"odinger-invariance of the three-point response function. 
Therefore, it might become unnecessary to request the co-variance of the correlation function,
which potentially could lead to problematic results \cite[app. B]{Henk25c}. 
The second equation (\ref{exposants}) will be needed shortly when comparing to explicit model results. 

To make the following discussion more explicit, we now make the ansatz of a pure power-law form $\Psi(Y)=\Psi_{\infty} Y^{\nu}$. 
Since Schr\"odinger-invariance leave the scaling function $\Phi$, and by implication $\Psi$, un-determined \cite{Henk94}, 
the ansatz is best considered as a kind of mathematical experiment. It will turn out to be satisfied in the two explicit examples studied below. 
Provided that is admissible, the integral in (\ref{3.4}) simplifies into, with $t=ys$ 
\begin{subequations} \label{gl:autoC}
\begin{align}
C(ys,s) &= C_{\infty}\, s^{-b}\, y^{\xi+d/2-2\wit{\delta}_2} \big(y+1\bigr)^{-d/2-\nu} \bigl(y-1\bigr)^{-2(\delta-\wit{\delta}_2)+\nu} \nonumber \\
&  ~~\times F_1\left( 2\wit{\xi}_2+1,2\wit{\delta}_2-d/2,d/2+\nu;2+d/2+2\wit{\xi}_2-2\wit{\delta}_2;\frac{1}{y},\frac{2}{y+1}\right)
\end{align}
where $F_1$ is one of the Appell functions \cite{Prud3} of which some identities are collected in the appendix. We also defined as a short-hand
\BEQ
C_{\infty} := T \Psi_{\infty} \frac{\Gamma(2\wit{\xi}_2+1)\,\Gamma(1+d/2-2\wit{\delta}_2)}{\Gamma(2+d/2+2\wit{\xi}_2-2\wit{\delta}_2)}
\EEQ
\end{subequations}

\subsection{Single-time correlator} 

We return to (\ref{corr-gen}) and set $s=t-\vep$ where we shall take the $\vep\to 0$ limit as soon as possible. 
Following \cite{Duva24}, by making a translation $\vec{R}\mapsto \vec{R}-\demi \vec{r}$ 
in the spatial integration variable, and also rescaling in the temporal integration variable, we find 
\begin{subequations} \label{gl:corr-spat} 
\begin{align}
C(t;r) &= T t^{2\xi} \int_0^{t} \!\D u\: u^{2\wit{\xi}_2}\, \bigl(t-u\bigr)^{-4\wit{\delta}_2} 
\int_{\mathbb{R}^d} \!\D\vec{R}\: \exp\left[ -\frac{{\cal M}}{t-u}\left( \frac{\vec{r}^2}{4} + \vec{R}^2\right)\right]   \nonumber \\
& ~~\times  \label{3.7}
\Phi\left( \frac{\vec{r}^2}{\vep} \right) \vep^{-2(\delta-\wit{\delta}_2)} \\
&= T \Phi_{\infty}\, t^{2\xi} \int_0^{t} \!\D u\: u^{2\wit{\xi}_2}\, \bigl(t-u\bigr)^{-4\wit{\delta}_2} \exp\left[ -\frac{{\cal M}}{4}\frac{\vec{r}^2}{t-u}\right] 
\int_{\mathbb{R}^d} \!\D\vec{R}\: \exp\left[ -\frac{{\cal M}}{2} \frac{\vec{R}^2}{t-u} \right]  ~~\label{3.8} \\
&= \mathscr{C}_{\infty}\, t^{-\mathfrak{b}} \int_0^1 \!\D v\: \bigl(1-v\bigr)^{2\wit{\xi}_2} v^{d/2-4\wit{\delta}_2} \exp\left[ -\frac{{\cal M}}{4} \frac{\vec{r}^2}{t} \frac{1}{v} \right]
\label{3.9} \\
&= \mathscr{C}_{\infty}\, \Gamma(4\wit{\delta}_2-d/2)\, t^{-\mathfrak{b}}\, e^{-\frac{{\cal M}}{4} \frac{\vec{r}^2}{t}}\: 
    U\left( 2\wit{\xi}_2+1,4\wit{\delta}_2-d/2;\frac{{\cal M}}{4} \frac{\vec{r}^2}{t} \right) 
\label{3.10}
\end{align}
\end{subequations}
where we abbreviated
\BEQ \label{gl:expo-spat}
\mathfrak{b} = -2\xi-2\wit{\xi}_2+4\wit{\delta}_2 -d/2-1 \;\; ,  \;\;
\mathscr{C}_{\infty} = T \Phi_{\infty} \left( \frac{2\pi}{{\cal M}}\right)^{d/2}
\EEQ
When going from (\ref{3.7}) to (\ref{3.8}) we take the limit $\vep\to 0$ inside the scaling function $\Phi$ which reduces to a constant $\Phi_{\infty}$. 
Taking this limit also produces\footnote{This further simplifies the two-time auto-correlator (\ref{3.4},\ref{gl:autoC}) and gives $b=\mathfrak{b}$.} the constraint $\wit{\delta}_2=\delta$.
The integral over $\vec{R}$ in (\ref{3.8}) is found trivially. $U$ is the Kummer/Tricomi hypergeometric function \cite{Prud3}. 

\subsection{Comparison with models} 

We started from (\ref{corrFT}) which is valid whenever the thermal bath white noise gives the relevant contribution. Now, we compare (\ref{3.4},\ref{gl:autoC}) 
and (\ref{gl:corr-spat}) to two models with $\mathpzc{z}=2$ and where noisy initial conditions are irrelevant: 
(i) the $1D$ Glauber-Ising model quenched to its critical point at $T=0$ \cite{Bray97,Godr00a,Henk04a,Henk25b} and (ii) the spherical model quenched onto its
critical point $T=T_c>0$ in $d>2$ dimensions \cite{Godr00b}. 

\subsubsection{Example 1: $1D$ Glauber-Ising model}

\begin{figure}[tb]  
\begin{center}
\includegraphics[width=0.5\hsize]{stoimenov15-autoC-IEL.eps}
\end{center}
\caption[fig3]{Auto-correlator scaling function $C(ys,s)=f_C(y)$ according to (\ref{gl:autoC-GI1D}) with $C_{\infty}=\frac{\sqrt{8\,}}{\pi}$ and for several values of $\nu$. 
The exact solution (\ref{1DGIC}) corresponds to $\nu=0$. \label{fig3} }
\end{figure}

The exactly known two-time auto-response function is \cite{Godr00a,Lipp00} 
\BEQ
R_{\rm GI}(t,s;r) = s^{-1} \left( \frac{t}{s} - 1 \right)^{-1/2} \exp\left[ - \frac{1}{2}\frac{r^2}{t-s}\right] 
\EEQ
This agrees with the expected form (\ref{2reponse}) and we read off $\delta=\frac{1}{4}$, $\xi=0$ and $\wit{\xi}=-\demi$ (we need not fix the non-universal constant $\cal M$). 
The exact two-time auto-correlator is \cite{Godr00a,Lipp00}
\BEQ \label{1DGIC}
C_{\rm GI}(ys,s) = \frac{2}{\pi} \arctan\sqrt{\frac{2}{y-1}\,}\: = \frac{2}{\pi} \arcsin \sqrt{\frac{2}{y+1}\,}
\EEQ
This means that the ageing exponent $b=0$ which from the second relation (\ref{exposants}) leads to $2\wit{\delta}_2 = 2\wit{\xi}_2+1$. A further constraint follows
since (\ref{1DGIC}) is finite for $y=1$. From (\ref{gl:autoC}) this gives the additional condition $2\wit{\xi}_2+\demi+\nu=0$ such that the form of the auto-correlator
\BEQ \label{gl:autoC-GI1D}
C(t,s) = C_{\infty}\, y^{\nu} \bigl(y+1\bigr)^{-\demi-\nu} 
F_1\left(\demi-\nu,-\nu,\demi+\nu;\frac{3}{2};\frac{1}{y},\frac{2}{y+1}\right)
\EEQ
depends on the single parameter $\nu$. This is shown in figure~\ref{fig3} for several values of $\nu$. The correct $y\gg 1$ asymptotics is matched when $C_{\infty}=\frac{2\sqrt{2\,}}{\pi}$.
Although the asymptotic behaviour for $y\gg 1$ is $\nu$-independent, strong differences occur in the opposite case $y\to 1^{+}$. 
For the Ising model, it obviously follows from the discrete nature of the Ising spins $\sigma=\pm 1$, that $C(s,s)=1$. This last constraint leads from
(\ref{gl:autoC-GI1D}) to the condition $C(s,s)\stackrel{!}{=} 1 = \frac{1}{\sqrt{\pi\,}} \frac{\Gamma(\demi+\nu)}{2^{\nu}\Gamma(1+\nu)}$ 
which has a singularity at $\nu=-\demi$. For $\nu>-\demi$ the only solution is $\nu=0$, see also \cite{Henk06}. 
Figure~\ref{fig3} shows that this indeed reproduces the exact solution (\ref{1DGIC}), also confirmed by the identity 
$\bigl(y+1\bigr)^{-\demi}F_1\bigl(\demi,0,\demi;\frac{3}{2};\frac{1}{y},\frac{2}{y+1}\bigr)=\frac{1}{\sqrt{2\,}}\arcsin\sqrt{\frac{2}{y+1}\,}$ using the appendix, as it should. 
We still notice the end values $\wit{\xi}_2=-\frac{1}{4}$ and $\wit{\delta}_2=\delta=\frac{1}{4}$. 

We now turn to the single-time temporal-spatial correlator for which the exactly known result is \cite{Bray97,Godr00a} 
\BEQ
C_{\rm GI}(t;r) = \erfc\left( \frac{r}{2\sqrt{t\,}} \right)
\EEQ
with the complementary error function, to be compared with (\ref{gl:corr-spat}). With the known values of the exponents, we confirm in (\ref{gl:expo-spat}) that $\mathfrak{b}=0$, 
as it should be. Inserting into (\ref{3.10}) 
we find with \cite[(7.11.4.14)]{Prud3}
\BEQ \label{gl:Ising1D}
C(t;r) 
= \mathscr{C}_{\infty} \sqrt{\pi\,}\, e^{-\frac{{\cal M}}{4}\frac{r^2}{t}}\: U\left(\demi,\demi;\frac{{\cal M}}{4}\frac{r^2}{t}\right) 
= \mathscr{C}_{\infty} \pi\: \erfc\left( {{\cal M}}^{1/2} \frac{r}{2\sqrt{t\,}} \right)
\EEQ
and it only remains to fix the overall normalisations correctly. 

\subsubsection{Example 2: critical spherical model} 

The exact two-time response function is \cite{Godr00b,Pico02} 
\BEA \label{3.14}
R_{\rm SM}(ys,s;r) = \frac{1}{(4\pi)^{d/2}} \frac{y^{-\digamma/2}}{\bigl( y-1\bigr)^{d/2}}  \exp\left[ -\frac{1}{2} \frac{r^2}{t-s}\right] \;\; , \;\;
\digamma = \left\{ \begin{array}{ll} d/2-2 & \mbox{\rm ~;~ if $2 < d <4$} \\
                                     0     & \mbox{\rm ~;~ if $d\geq 4$} 
                  \end{array} \right. ~~
\EEA
and comparing with (\ref{2reponse}) we read off $\xi = - \wit{\xi}=-\digamma/2$ and $\delta=d/4$ but do not fix the non-universal constant $\cal M$. 
The exact two-time auto-correlator is given by \cite{Godr00b,Pico02}
\BEA
C_{\rm SM}(t,s) &=& (4\pi)^{-d/2+1} 2T_c\, s^{-(d/2-1)}\, y^{-\digamma/2} \int_0^1 \!\D w\: w^{\digamma}\, \bigl( y+1-2w\bigr)^{-d/2}
\\
&=& \frac{2T_c}{(4\pi)^{d/2}} \frac{\Gamma(\digamma+1)}{\Gamma(\digamma+2)}\, s^{-(d/2-1)}\, y^{-\digamma/2} \left( \frac{2}{y+1}\right)^{d/2}
{}_2F_1\left( \frac{d}{2},\digamma+1;\digamma+2;\frac{2}{y+1}\right) \nonumber 
\EEA
Comparison with (\ref{3.4}), and also assuming $\Psi(Y)=\Psi_{\infty}$, shows that $d/2-2\wit{\delta}_2=0$ such that we can simplify (\ref{gl:autoC}) and we then have
\BEA
C(t,s) &=& T_c \Psi_{\infty} s^{-{b}}\, y^{\xi} \int_0^1 \!\D w\: w^{2\wit{\xi}_2} \bigl( y+1 -2w\bigr)^{-d/2}
\nonumber \\
&=& \mathscr{C}_{\infty}\, s^{-b}\, y^{\xi}\bigl(y+1\bigr)^{-d/2-\nu}\bigl(y-1\bigr)^{-\nu} 
{}_2F_1\left( \frac{d}{2}+\nu, 2\wit{\xi}_2+1;2\wit{\xi}_2+2;\frac{2}{y+1}\right)
\EEA
so that we identify $2\wit{\xi}_2=\digamma$. This gives $\delta=\wit{\delta}_2=d/4$ and $\wit{\xi}_2=\frac{d}{4}-1$ for $2<d<4$, or $\wit{\xi}_2=0$ for $d\geq 4$, 
which when inserted in the second relation (\ref{exposants}) does produce $b=d/2-1$ as expected.

The exact single-time correlator is, up to overall and dimension-dependent normalisation constants \cite{Godr00b} 
\BEA \label{gl:corr-SM}
C_{\rm SM}(t;\vec{r}) &=& \mathscr{C}_{\infty}^{(1)} T_c\, t^{1-d/2} \int_0^1 \!\D v\: (1-v)^{\digamma} v^{-d/2}\, e^{-\frac{1}{4}\frac{r^2}{t}\frac{1}{v}} 
\nonumber \\
&=& \mathscr{C}_{\infty}^{(2)} T_c\, t^{1-d/2}\, e^{-\frac{1}{4}\frac{r^2}{t}}\, U\left( \digamma+1,\frac{d}{2}; \frac{1}{4} \frac{r^2}{t}\right) 
\nonumber \\
&=& \mathscr{C}_{\infty}^{(3)} T_c\, |\vec{r}|^{2-d} 
\left\{ \begin{array}{ll}  \exp\left[ -\frac{1}{4} \frac{\vec{r}^2}{t} \right]                     & \mbox{\rm ~;~~ if $2<d<4$} \\[0.25cm]
                           \int_{\frac{1}{4} \frac{\vec{r}^2}{t}}^{\infty} \D z\: z^{d/2-2} e^{-z} & \mbox{\rm ~;~~ if $4<d$} 
        \end{array} \right. \label{gl:spherique}
\EEA
and the last line with more explicit expressions follows from $\digamma$ as given in (\ref{3.14}). For $2<d<4$ we also used the identity $U(a-1,a;z)=z^{1-a}$ where $a$ is not a 
positive integer. 
For comparison with (\ref{gl:corr-spat}), we recall first the generic prediction (with normalisation ${\cal C}_{\infty}$) 
\BEQ
C(t;r) = {\cal C}_{\infty}\, t^{-b}\, e^{-\frac{\cal M}{4}\frac{r^2}{t}} U\left(2\wit{\xi}_2+1,4\wit{\delta}_2-\frac{d}{2};\frac{\cal M}{4} \frac{r^2}{t}\right) 
\EEQ
and we can read off once more the universal exponents $\delta=\wit{\delta}_2=d/4$ and $\wit{\xi}_2=-\xi = \digamma/2$. 
Inserting into (\ref{gl:expo-spat}) reproduces
the expected $\mathfrak{b}=-1+d/2=b$. Hence, with the same values of for the two-time auto-correlator, we reproduce the prediction (\ref{3.10}), upon the
correct identification of the non-universal normalisations. The exponent values are gathered in table~\ref{tab:1}.

\section{Conclusions} \label{sec:4}

\begin{table}  
\begin{center}
\begin{tabular}{|lrr|ccccc|}  \hline
\multicolumn{3}{|l|}{~} & \multicolumn{5}{r|}{~} \\[-0.4cm] 
\multicolumn{3}{|l|}{model}              & ~$\delta=\wit{\delta}$~ & ~$\xi$~  & ~$\wit{\xi}$~ & ~$\wit{\delta}_2$~ & ~$\wit{\xi}_2$~ \\ \hline
Glauber-Ising & ~$T=0$~~    & ~$d=1$~    & $1/4$                   & $0$      & $-1/2$        & $1/4$              & $-1/4$          \\
spherical     & ~$T=T_c$~   & ~$2<d<4$~  & $d/4$                   & $1-d/4$  & $d/4-1$       & $d/4$              & $d/4-1$         \\
spherical     & ~$T=T_c$~   & ~$4<d$~    & $d/4$                   & $0$      & $0$           & $d/4$              & $0$             \\ \hline 
\end{tabular}\end{center}
\caption[tab1]{Values of the scaling dimensions needed in the calculation of correlators in exactly solved models.  
\label{tab:1}
}
\end{table}

Dynamical scaling is one of the central ingredients of physical ageing \cite{Stru78,Bray94a,Godr02,Cugl03,Cugl15} 
but it has ever been elusive to fix the from of the universal scaling functions of single-time or two-time correlators,
see figure~\ref{fig2}bd, in spite of  these being the quantities most easily measured in simulations or experiments. 
The most simple generalisation of dynamical scaling occurs when the dynamical exponent $\mathpzc{z}=2$ and is based on the 
Schr\"odinger Lie algebra, see \cite{Duva24} and references therein, which makes  Schr\"odinger-invariance a natural candidate for the determination of the scaling functions.
In contrast to our previous study \cite{Henk25c} where we assumed  that {\em both} responses and correlators transform covariantly, 
we admit here merely the covariant transformation of two-point and
three-point response functions and then reconstruct two-point correlators from (\ref{corrFT}), valid for model-A-type dynamics. 
Here we studied the peculiar case when the relevant noise comes from the thermal bath 
{(modelled by a gaussian white noise of temperature $T_c$)}, 
{as it is the case after a quench onto $T=T_c$, but we also restricted ourselves to 
the small sub-set of systems where $\mathpzc{z}=2$.} Only {then} the use of (\ref{corrFT}) is admissible. 
Our input data were the known scaling exponents $\delta=\wit{\delta}$, $\xi$ and $\wit{\xi}$, 
see table~\ref{tab:1}, of the exact two-time response functions $R(t,s;r)$. 
In addition, we could side-step the more difficult question what specific form of the initial correlators should be used. 
Our results, which include the exponent equality $\lambda_C=\lambda_R$, see eq.~(\ref{gl:lambda}), supports our main idea: 
{\it the Schr\"odinger-invariant three-point response functions}, 
which for their calculation require the validity of time-translation-invariance and are therefore {\bf at} equilibrium, 
{\it can be adapted to non-equilibrium ageing via the rule (\ref{reponseHE})} \cite{Henk25c}. 
This leads to the explicit expressions (\ref{3.4},\ref{gl:autoC}) for $C(t,s)$ and (\ref{gl:corr-spat}) for $C(t;r)$, 
respectively; and with the additional constraint $\wit{\delta}_2=\delta$. 

We tested the idea explicitly in two exactly solvable models with $\mathpzc{z}=2$ and where the bath disorder is relevant: 
namely (i) the $1D$ Glauber-Ising model quenched to $T=0$ and (ii) the spherical model quenched onto $T=T_c$ for any $d>2$. 
The values of the scaling dimensions needed are collected in table~\ref{tab:1}. 
{Different values of the exponents correspond to distinct dynamical universality classes.}
The composite response operator $\wit{\phi^2}$ arising in the three-point function (\ref{corrFT}) has non-trivial scaling dimensions, notably $\wit{\xi}_2$. 
The success of these comparisons now calls for the study of models of phase-ordering, after a quench to $T<T_c$, 
{where naturally $\mathpzc{z}=2$ \cite{Bray94b} and to take the noisy initial state into account. 
In contrast to our starting point (\ref{corrFT}), it will then be required to analyse
four-point response functions $\bigl\langle \phi\phi\wit{\phi}\wit{\phi}\bigr\rangle_0$ \cite{Henk25e}. 
Schr\"odinger-invariance permits first to re-derive the generic properties of the
two-time correlators and responses as discussed before \cite{Henk25c} and second, 
to include a discussion of the single-time correlator $C(t;r)$ and the applicability of {\sc Porod}'s law \cite{Poro51,Bray94a}. 
However, and in contrast with what we found in the present work valid at $T=T_c$, 
Schr\"odinger-invariance alone is no longer able, for $T<T_c$, to provide explicit predictions for the form of the scaling functions $F_{C,R}$. 
It follows that generalisations of that Lie algebra must be sought,
possibly along the lines of \cite{Henk04b,Lorenz07a}, which is work in progress.} 

It appears significant that on one hand our results for the temporal-spatial single-time correlator 
include (\ref{gl:Ising1D}) which leads to sharp {and non-gaussian} Ising-like interfaces as predicted from Porod's law \cite{Poro51} and on the other hand
also permits soft interfaces as they occur in the spherical model (\ref{gl:spherique}), 
{which is of a gaussian nature although the exponents for $d<d^*$ are not in the mean-field universality class.} 
These cases are distinguished via the values of the scaling dimensions, 
viz. $\delta,\xi,\wit{\xi},\wit{\xi}_2$. 
{Studies in more general models will hopefully shed more light on the validity of the predictions for the single-time correlator
$C(t;r)$ and the two-time auto-correlator $C(t,s)$.}\footnote{{After this work was submitted, we derived the Schr\"odinger-invariance of the the 
{\em voter model} \cite{Ligg85,Tome01,Krap10,Corb24e,Godr24} 
in $d>0$ dimensions which turns out to be another example of a non-equilibrium critical system with $\mathpzc{z}=2$ \cite{Henk25f}.
Further exactly solved examples are listed in \cite{Henk25g}.}}

As a preparation for this, we considered here how to find the form of the correlators in a few situations, with $\mathpzc{z}=2$, 
but where the relevant noise comes from the thermal bath so that we could use 
(\ref{corrFT}) as our starting point. Our input data were the known scaling exponents $\delta=\wit{\delta}$, $\xi$ and $\wit{\xi}$, 
see table~\ref{tab:1}, of the exact two-time response functions $R(t,s;r)$. 
In addition, we could side-step the more difficult question what specific form of the initial correlators should be used. 
Our present work, which in particular allows to re-derive the anticipated exponent equality $\lambda_C=\lambda_R$, see eq.~(\ref{gl:lambda}), does confirm our main idea: 
{\it the Schr\"odinger-invariant three-point response functions}, 
which for their calculation require the validity of time-translation-invariance and are therefore {\bf at} equilibrium, 
{\it can be adapted to non-equilibrium ageing via the rule (\ref{reponseHE})} \cite{Henk25c}. 
This leads to the explicit expressions (\ref{3.4},\ref{gl:autoC}) for $C(t,s)$ and (\ref{gl:corr-spat}) for $C(t;r)$, 
respectively; and with the additional constraint $\wit{\delta}_2=\delta$. 

We tested the idea explicitly in two exactly solvable models with $\mathpzc{z}=2$ and where the bath disorder is relevant: 
namely (i) the $1D$ Glauber-Ising model quenched to $T=0$ and (ii) the spherical model quenched onto $T=T_c$ for any $d>2$. 
The values of the scaling dimensions needed are collected in table~\ref{tab:1}. 
The composite response operator $\wit{\phi^2}$ arising in the three-point function (\ref{corrFT}) has non-trivial scaling dimensions, notably $\wit{\xi}_2$. 
The success of these comparisons now calls for the study of models of phase-ordering, after a quench to $T<T_c$, 
where $\mathpzc{z}=2$ and to take the noisy initial state into account. 
Work along these lines is in progress. 

It appears significant that on one hand our results for the temporal-spatial single-time correlator 
include (\ref{gl:Ising1D}) which leads to sharp Ising-like interfaces as predicted from Porod's law and on the other hand
also permits soft interfaces as they occur in the spherical model (\ref{gl:spherique}). These cases are distinguished via the values of the scaling dimensions, 
viz. $\delta,\xi,\wit{\xi},\wit{\xi}_2$. 

\noindent
{\bf Acknowledgements:}  
This work was supported by the french ANR-PRME UNIOPEN (ANR-22-CE30-0004-01) and by PHC RILA (Dossier 51305UC).

\newpage 

\appsektion{Appell function identities}

We collect some identities on the Appell function $F_1$ which are needed in the main text. 
It is a hypergeometric function of two  variables, convergent for $|w|<1$ and $|z|<1$,  with the integral representation \cite[(7.2.4.42)]{Prud3}
\BEA
F_1\bigl(a,b_1,b_2;c;w,z\bigr) &=& F_1\bigl(a,b_2,b_1;c;z,w\bigr) = 
\sum_{k=0}^{\infty} \sum_{\ell=0}^{\infty} \frac{(a)_{k+\ell} (b_1)_k (b_2)_{\ell}}{(c)_{k+\ell}} \frac{w^k}{k!} \frac{z^{\ell}}{\ell!} \label{A1} \\
&=& \frac{\Gamma(c)}{\Gamma(a)\Gamma(c-a)} \int_0^1 \!\D u\: \frac{u^{a-1} \bigl(1-u\bigr)^{c-a-1}}{\bigl(1-u w\bigr)^{b_1} \bigl(1- u z\bigr)^{b_2}} \label{A2}
\EEA
and the Pochhammer symbol $(a)_k = \frac{\Gamma(a+k)}{\Gamma(a)}$. The series definition (\ref{A1}) implies 
\BEQ \label{A3} 
F_1\bigl(a,b,0;c;w,z) = {}_2F_1\bigl(a,b;c;w)
\EEQ
and one also has
\BEA
F_1\bigl(a,b_1,b_2;c;w,1\bigr)   &=& {}_2F_1\bigl(a,b_2;c;1\bigr) {}_2F_1\bigl(a,b_1;c-b_2;w) \:=\: \frac{\Gamma(c)\Gamma(c-a-b_2)}{\Gamma(c-a)\Gamma(c-b_2)} {}_2F_1\bigl(a,b_1;c-b_2;w) 
\nonumber \\ 
&& \label{A4} \\
F_1\bigl(a,b_1,b_2;c;1,1\bigr)   &=& {}_2F_1\bigl(a,b_1+b_2;c;1\bigr) \:=\: \frac{\Gamma(c)\Gamma(c-a-b_1-b_2)}{\Gamma(c-a)\Gamma(c-b_1-b_2)} 
\label{A5} \\
F_1\bigl(a,b_1,b_2;c;-1,-1\bigr) &=& {}_2F_1\bigl(a,b_1+b_2;c;-1\bigr)  
\label{A6}
\EEA
See \cite{Prud3} for values of ${}_2F_1\bigl(a,b;c;-1\bigr)$.


\end{document}